\documentclass[12pt]{article}

\usepackage[super]{cite}
\usepackage{fullpage}

\usepackage{amssymb}
\usepackage{amsmath}  
\usepackage{times}  
\usepackage[urlcolor=blue]{hyperref}
\usepackage[pdftex]{color}
\usepackage{graphicx}

\usepackage{placeins}

\newcommand{\spacing}[1]{\renewcommand{\baselinestretch}{#1}\large\normalsize}
\spacing{1.1}

\newenvironment{affiliations}{%
    \setcounter{enumi}{1}%
    \setlength{\parindent}{0in}%
    \slshape\sloppy%
    \begin{list}{\upshape$^{\arabic{enumi}}$}{%
        \usecounter{enumi}%
        \setlength{\leftmargin}{0in}%
        \setlength{\topsep}{0in}%
        \setlength{\labelsep}{0in}%
        \setlength{\labelwidth}{0in}%
        \setlength{\listparindent}{0in}%
        \setlength{\itemsep}{0ex}%
        \setlength{\parsep}{0in}%
        }
    }{\end{list}\par\vspace{12pt}}

\renewenvironment{abstract}{%
    \setlength{\parindent}{0in}%
    \setlength{\parskip}{0in}%
    \bfseries%
    }{\par\vspace{-6pt}}

\newenvironment{methods}{%
    \section*{\large Methods}%
    \setlength{\parskip}{12pt}%
    }{}

\newenvironment{addendum}{%
    \setlength{\parindent}{0in}%
    \small%
    \begin{list}{Acknowledgements}{%
        \setlength{\leftmargin}{0in}%
        \setlength{\listparindent}{0in}%
        \setlength{\labelsep}{0em}%
        \setlength{\labelwidth}{0in}%
        \setlength{\itemsep}{12pt}%
        }
    }
    {\end{list}\normalsize}

\newcommand*{\addendumlabel}[1]{\textbf{#1}\hspace{1em}}

\newcommand{\blue}{\color{blue}}

\newcommand{\bnabla}{\mbox{\boldmath $\nabla$}}

\title{Weak magnetic field changes over the Pacific due to high conductance in lowermost mantle}

\author{Mathieu Dumberry$^{1}$ \& Colin More$^{1}$}

\begin{document} 

\maketitle

\begin{affiliations}
 \item Department of Physics, University of Alberta, Edmonton, T6G 2E1, Canada
\end{affiliations}


\baselineskip20pt

\begin{abstract}
For the past few centuries, the temporal variation in the Earth's magnetic field  in the Pacific region has been anomalously low.  The reason for this is tied to large scale flows in the liquid outer core near the core-mantle boundary, which are weaker under the Pacific and feature a planetary scale gyre that is eccentric and broadly avoids this region.   However, what regulates this type of flow morphology is unknown.  Here, we present results from a numerical model of the dynamics in Earth's core that includes electromagnetic coupling with a non-uniform conducting layer at the base of the mantle. We show that when the conductance of this layer is higher under the Pacific than elsewhere, the larger electromagnetic drag force weakens the local core flows and deflects the flow of the planetary gyre away from the Pacific.  The nature of the lowermost mantle conductance remains unclear, but stratified core fluid trapped within topographic undulations of the core-mantle boundary is a possible explanation.  
\end{abstract}

The Earth's magnetic field is generated by electrical currents flowing within its conducting iron core. These currents, in turn, are driven and maintained against decay by motions in the fluid outer core, likely convective in nature\cite{jones15}.  Core flows produce changes in the magnetic field, including those observed at the Earth's surface, a temporal fluctuation which is referred to as secular variation (SV)\cite{jackson15}.  Past observations can be used to reconstruct how  the magnetic field and its SV have changed with time\cite{jackson15,bloxham85,jackson00}.  Fig. 1a shows the mean intensity of the radial component of the SV at the core-mantle boundary (CMB) over the time period 1590-1990 from the model in ref. (\citen{jackson00}).  The SV is not uniformly distributed and is distinctly weaker under a broad region of the Pacific ocean, a pattern that was first noticed in the 1930's\cite{fisk31,fleming39}.   Satellite observations in the past few decades have confirmed the low geomagnetic SV in the Pacific\cite{finlay12}, thus alleviating a concern that models based on earlier observations might include biases related to geographic coverage.  Fig. 1b shows the SV of the radial component of the field in 2015 from the model of ref. (\citen{finlay16}).

An electromagnetic (EM) screening effect, whereby a larger mantle conductance (the integral of the electrical conductivity, $\sigma$, over a thickness $\Delta$) above the CMB in the Pacific region filters a greater part of the SV, was originally proposed as a possible explanation\cite{cox62,runcorn92}.  However, the required conductance (see Supplementary Information) is too high by two orders of magnitude than modern estimates inferred from Earth's nutations\cite{buffett92,buffett02,koot13}, length-of-day variations\cite{holme98b}, and attenuation of Alfv\'en waves\cite{schaeffer16b}, which collectively point to a globally averaged conductance in the range of $1-3\times10^8$ S.  The strength of the radial magnetic field is lower in the Pacific (Supplementary Information, Figure S3), but not sufficiently to explain the level of SV decrease, so the explanation must also involve the geographic pattern of core flows\cite{vestine66,doell71}.  Flow maps at the CMB can be reconstructed on the basis of the observed SV; Fig. 2 shows the time-averaged flow between 1940-2010 taken from the model of ref. (\citen{gillet15}).  The dominant structure is a westward, eccentric planetary gyre\cite{finlay12,gillet15,pais08,aubert13,aubert14,aubert20}, flowing in the equatorial and mid-latitude regions of the Atlantic hemisphere but deflected to polar latitudes when entering the Pacific hemisphere.  

A first-order explanation for the lower SV in the Pacific is then simply that the mean planetary gyre, which features the fastest flows, broadly avoids this region\cite{finlay12}.  Flows are not altogether absent in the Pacific, but tend to be weaker and more transitory.   To give a quantitative measure of these weaker flows, we define the Pacific region as lying within a spherical cap with a polar angle of 62$^\circ$, centred on the equator at longitude 180$^\circ$E.  We have computed maps of the flow for each year between 1940-2010 from the flow model of ref. (\citen{gillet15}) shown in Fig. 2.  The mean root-mean-square (r.m.s.) flow amplitude over the Pacific for this time period is ${\cal V}_p=6.868$ km/yr, whereas the mean r.m.s. global flow amplitude is ${\cal V}$ = 10.472 km/yr, for a ratio between the two of ${\cal R}_v = {\cal V}_p/{\cal V} = 0.656$.  The minimum and maximum annual ratios over the time period covered by the model are 0.597 and 0.744, respectively.  

But why are core flows weaker in the Pacific and why is this region avoided by the gyre? This present-day geometry  may simply reflect a transient configuration maintained over a few centuries by the turbulent convective dynamics of the core\cite{schaeffer17}.  However, it may also be an imprint of a coupling between the core and a heterogeneous lowermost mantle\cite{runcorn56,doell71}.  Thermal coupling can reproduce some features of the large scale flows\cite{olson02,christensen03,aubert07}, though their amplitudes are too small by a factor of 10.  A gyre that resembles that of Fig. 2 results when a hemispherical pattern of inner core growth is further imposed\cite{aubert13b}, but it is unclear whether this could be maintained since convection within the inner core is unlikely\cite{labrosse14}.  Here, we show that a non-uniform EM coupling, larger over the Pacific, can explain the core flow morphology.

\section*{\large EM coupling in a quasi-geostrophic model of the fluid core}

To show this, we use a quasi-geostrophic model of magnetoconvection in the Earth's core\cite{more18}, modified to include EM coupling (see Methods). The latter acts as a drag force slowing down core flows. Regions of higher mantle conductance should exert a stronger EM drag force, reducing core flow speeds and leading to a lower SV.   EM coupling is parameterized in terms of a timescale $\tau_{em}$ which represents the characteristic timescale for a differentially moving core flow structure to fall back into co-rotation with the mantle.  To give a typical measure of $\tau_{em}$, a conductance of $10^8$ S should attenuate unforced core flows in approximately 30 yr (ref. \citen{dumberry08c}).  Time in our numerical model is scaled by the Alfv\'en timescale $\tau_A$, the time it takes for Alfv\'en waves to twice traverse the fluid core in the direction perpendicular to rotation.  For Earth's core, $\tau_A$ is approximately equal to 6 yr (ref. \citen{gillet10}), so a uniform conductance of the order of $10^8$ S corresponds to $\tau_{em}\approx5\cdot \tau_A$.  To keep a similar ratio of EM attenuation to Alfv\'en timescales as in Earth's core, we adopt $\tau_{em} = 5\cdot \tau_A$ at all points of flow contact with the CMB, except within a spherical cap in the Pacific region (see Methods), where we use a smaller EM coupling timescale (which we denote by $\tau^p_{em}$) to simulate the effect of a locally larger conductance.  It is convenient to introduce a quantity ${\cal X} = \tau_{em}/\tau^p_{em}$ which represents the factor of increase of EM coupling strength in the Pacific region versus elsewhere on the CMB.

We have carried out a suite of numerical experiments using four different values of the Rayleigh number ($Ra=2.5 \times 10^8$, $5 \times 10^8$, $7.5 \times 10^8$ and $10^9$) and two different values of the magnetic Prandtl number ($P_m=0.1$ and $0.01$) (see Methods for the definition of $Ra$ and $P_m$.)  For each combination of $Ra$ and $P_m$, we have varied ${\cal X}$ from 1 to 12.  Our quasi-geostrophic model is engineered to capture the decadal timescale flow dynamics in Earth's core, so we concentrate our analysis on how a non-uniform EM coupling at the CMB affects the flow structures.  (Additional analysis of the SV is presented in the Supplementary Information.)  Fig. 3a shows the axial vorticity and velocity for a numerical experiment with $Ra=5 \times 10^8$, $P_m=0.1$, and with $\tau^p_{em}$ set equal to $\tau_{em}$ (${\cal X} =1$). This represents a reference case where EM coupling is everywhere equal at the CMB and serves as a benchmark against which the effect of a larger EM coupling in the Pacific can be measured.  To be precise, the axial vorticity map is a snapshot in time, but the velocity map is an average over a time-window of $11\cdot\tau_A$, similar to that over which the flow map of Fig. 2 is time-averaged.  Hence, we can directly compare the flow structure in our model with the flow map derived from the observed SV in Fig. 2.  Although the mean circulation of our model is also westward, in contrast to the gyre structure of Fig. 2, the flow speed increases in a more monotonic manner with distance from the rotation axis.  Since the Pacific region is concentrated on the equator, its r.m.s. flow is larger than the global average (${\cal R}_v =1.152$). 

Fig. 3b-d show how the axial vorticity and time-averaged flow are modified when EM coupling in the Pacific is increased by a factor ${\cal X} = 3$, $5$ and $10$.  The vorticity maps highlight how individual convective columns are weakened by a stronger EM coupling. A similar pattern is observed on maps of the SV (Supplementary Information Figure S4).  The flow maps show how the morphology of the large scale circulation is altered: the larger ${\cal X}$ is, the more the mean westward flow in the Pacific hemisphere is squeezed towards the inner radial boundary in order to avoid the region of higher conductance.   Though differences exist between the mean westward planetary gyre in our QG model and that derived from the observed SV, a larger EM coupling over the Pacific leads to an eccentric gyre which avoids the equatorial region of the Pacific, as is seen in Fig. 2.   

The flow maps of Fig. 3 further illustrate that the r.m.s. flow amplitude in the Pacific (${\cal V}_p$) gets smaller than the global average (${\cal V}$) as the EM coupling strength in the Pacific is increased.   Fig. 4 shows how the ratio ${\cal R}_v = {\cal V}_p/{\cal V}$ decreases as a function of ${\cal X}$ for each of the 8 possible combinations of $Ra$ and $P_m$.  To match the ratio ${\cal R}_v=0.656$ derived from the observed SV, the conductance of the lowermost mantle in the Pacific must be approximately a factor 6 to 9 larger than elsewhere.  Since the Pacific cap occupies approximately a quarter of the surface area of the CMB, this translates into a conductance over the Pacific which is a factor of 2.7 to 3 larger than the global average. This factor is imprecise not only because of the range of ${\cal R}_v$ values that result from different choices of model parameters, but also because of the fluctuations about the mean value for each suite of models.  Pinning down the precise factor is further hampered by the fact that our model represents a simplified version of core dynamics and the morphology of the mean planetary gyre differs from that of Fig. 2.

\section*{\large The nature of the lowermost mantle conductance}

The question that remains to be addressed is the source of the lowermost mantle conductance and why it is larger over the Pacific.  Inferences from Earth's diurnal nutations require a global conductance of the order of $\sigma \cdot \Delta \approx 10^8$ S concentrated in a thin layer ($\Delta \approx 0.1 - 1$ km) of high electrical conductivity ($\sigma \approx 10^5-10^6$ S m$^{-1}$) at the base of the mantle\cite{buffett92,buffett02,koot13}.  The most abundant mantle mineral at CMB conditions is post-perovskite, but its electrical conductivity is $\sigma \sim2\times10^2$ S m$^{-1}$ (ref. \citen{ohta08}), which is too low.  Several ideas for how the lowermost mantle may be enriched in iron -- and thus be more electrically conducting -- have been proposed\cite{petford05,kanda06,otsuka12,dobson05,labrosse07}, but it remains unclear whether a conductivity approaching that of the core is possible (see Supplementary Information).  

An alternative explanation is that pockets of strongly stratified core fluid may be trapped by local cavities in CMB topography, remaining broadly stationary with respect to the mainstream core flows.  Radial magnetic field lines would be strongly anchored in these fluid pockets, leading to an efficient EM drag on flows underneath\cite{glane18}.  In terms of EM coupling, trapped fluid in CMB cavities could then mimic  a highly electrically conducting lower mantle, although this requires further testing.  The Pacific region may thus have a greater proportion of cavities, or deeper ones, than elsewhere on the CMB.  

\section*{\large Longer timescale SV and geodynamic implications}

The observed SV over the past few decades is controlled dominantly by the action of core flows on magnetic field structures.   Over longer timescales, changes in the large scale magnetic field that we see at Earth's surface should reflect instead a balance between dynamo action and magnetic diffusion within the core.  The upward diffusion of magnetic field structures to the top of the core should be largely unaffected by the pattern of CMB conductance.  Hence, the morphology of the SV over millennial timescales and longer is likely to be less impacted by a non-uniform CMB conductance than the decadal SV.  Though this remains to be demonstrated, this would explain why the millennial SV does not appear to be weaker in the Pacific region\cite{korte05b,lawrence06}.  

The westward flows over most of the CMB impart a westward EM torque on the mantle\cite{holme98b}. Flows in the Pacific, though weaker, tend to be eastward.  Hence, an additional consequence of a stronger EM coupling in the Pacific is that this local eastward EM torque can cancel a significant portion of the westward torque acting on the rest of the CMB\cite{holme00}.  This reduces the need for an additional eastward torque on the mantle, necessary to ensure its long term angular momentum balance \cite{buffett99}.  This torque has been suggested to be from gravitational coupling with the inner core\cite{buffett96,dumberry07b,pichon16}, and a reduction of its amplitude relaxes the requirement for a low inner core viscosity\cite{buffett97}.  

Although we can build a prediction of the SV from our QG model (see Supplementary Information), some dynamical feedbacks are absent. Hence, to further explore the consequences of a non-uniform EM coupling, the next step is to adopt a more complete, three-dimensional model of the core dynamics.  Not only this will permit a better match of the observed gyre\cite{aubert13,aubert14,aubert20}, this will also allow for a  more direct comparison between the predicted and observed SV at both decadal and millennial timescales.  Our estimate of the $6-9$ factor of conductance increase over the Pacific may be modified as a result.   Whether originating from mantle mineralogy or trapped fluid in CMB cavities, the conductance at the CMB is most likely not uniform.  Our results demonstrate the imprint that this may have on core flows and the SV.  Further investigating the effects of non-uniform EM coupling on core dynamics can thus help constrain the structures, composition and evolution of the lowermost mantle and the CMB region.  


\newpage

\FloatBarrier

\begin{figure}
\begin{center}
\includegraphics[width=14cm]{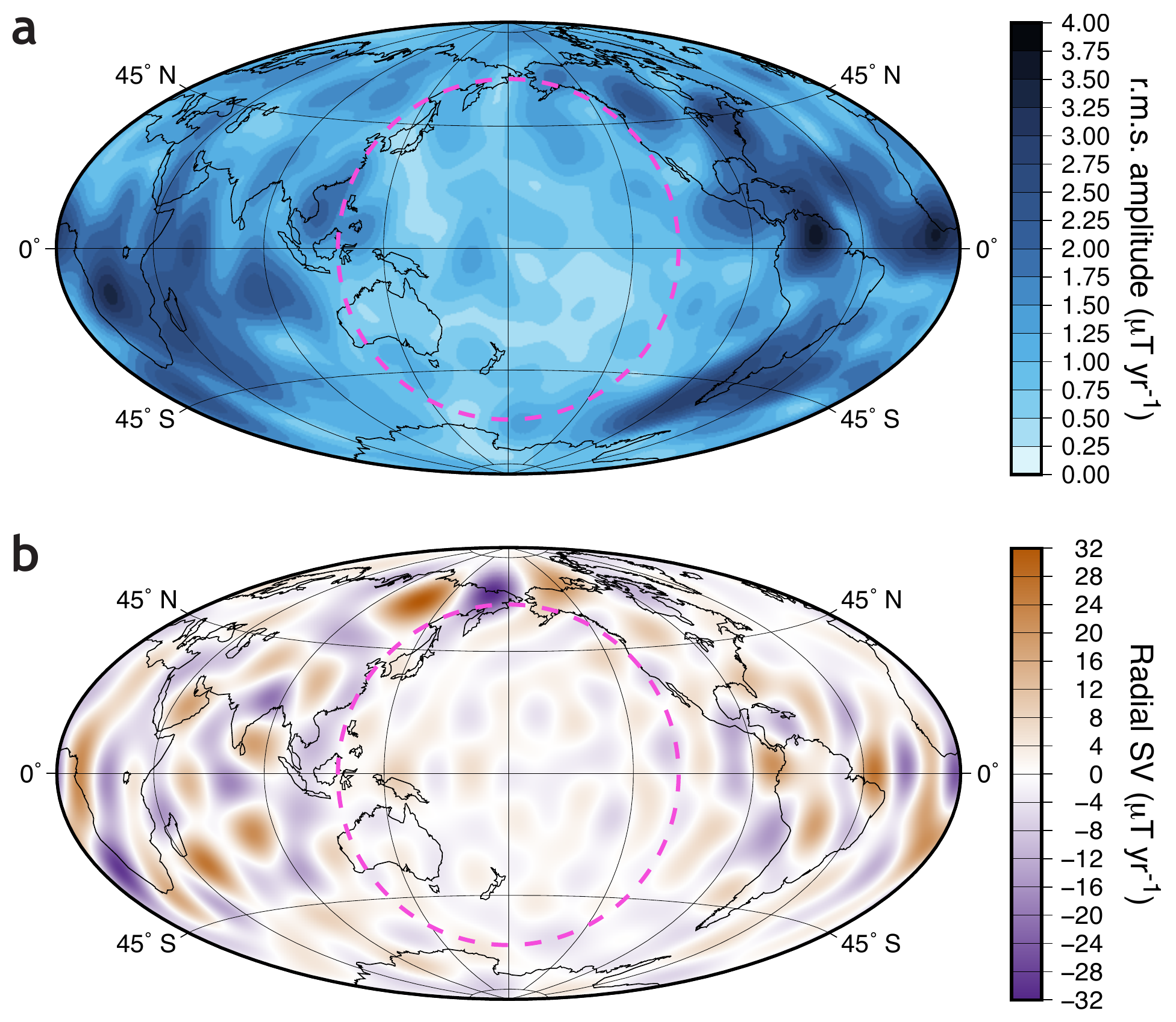}
\caption{ {\bf The low geomagnetic secular variation in the Pacific.}  {\bf a}, The mean intensity of the radial component of the SV at the CMB, $|\dot{B}_r|$, over the time-period $1590-1990$ from the {\em gufm} field model \cite{jackson00}.  The r.m.s. amplitude of $|\dot{B}_r|$ in the Pacific (pink dashed circle) is $796.32$ nT/yr, the global average is $1332.86$ nT/yr, for a ratio of $0.5975$. {\bf b}, The radial component of the SV at the CMB, $\dot{B}_r$, in $2015$ from the {\em CHAOS-6} field model \cite{finlay16} truncated at spherical harmonic degree $16$.  The r.m.s. amplitude of $|\dot{B}_r|$ over the Pacific (pink dashed circle) is $2034.13$ nT/yr, the global average is $4401.09$ nT/yr, for a ratio of $0.4622$.}
\end{center}
\end{figure}

\newpage

\FloatBarrier

\begin{figure}
\begin{center}
\includegraphics[width=9cm]{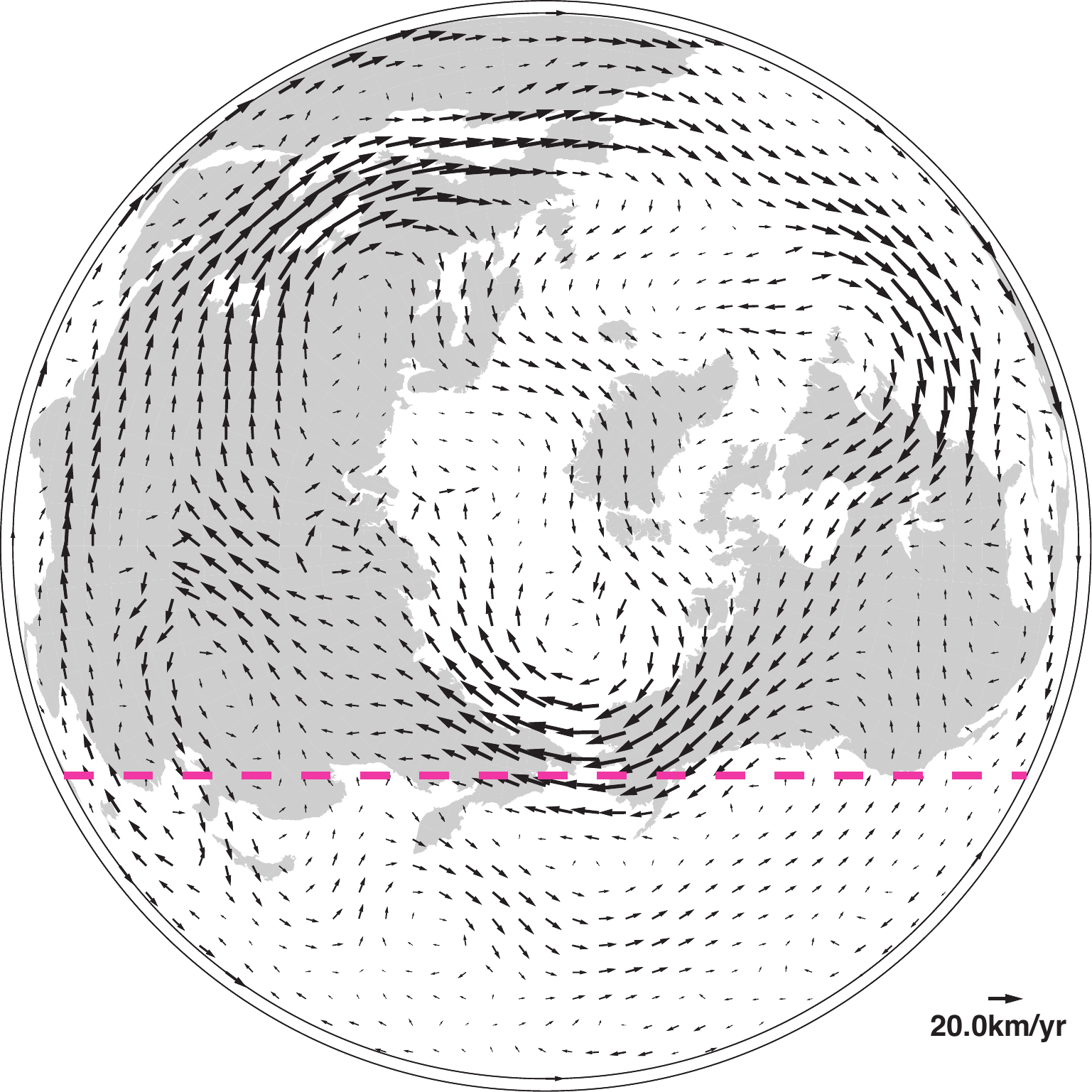} 
\caption{ {\bf The westward, eccentric planetary gyre in Earth's fluid core.}  The time-averaged core flow circulation at the CMB over the period 1940-2010 projected onto the equatorial plane inside the core (from the flow model of ref. (\citen{gillet15}), truncated at spherical harmonic degree 14).  The flow is assumed invariant in the direction of the rotation axis.  Northern hemisphere continents, projected onto the CMB, are shown for geographic reference.  The pink dashed line shows the boundary of our definition of the Pacific region.}
\end{center}
\end{figure}

\newpage 

\FloatBarrier

\begin{figure}
\begin{center}
\includegraphics[width=17cm]{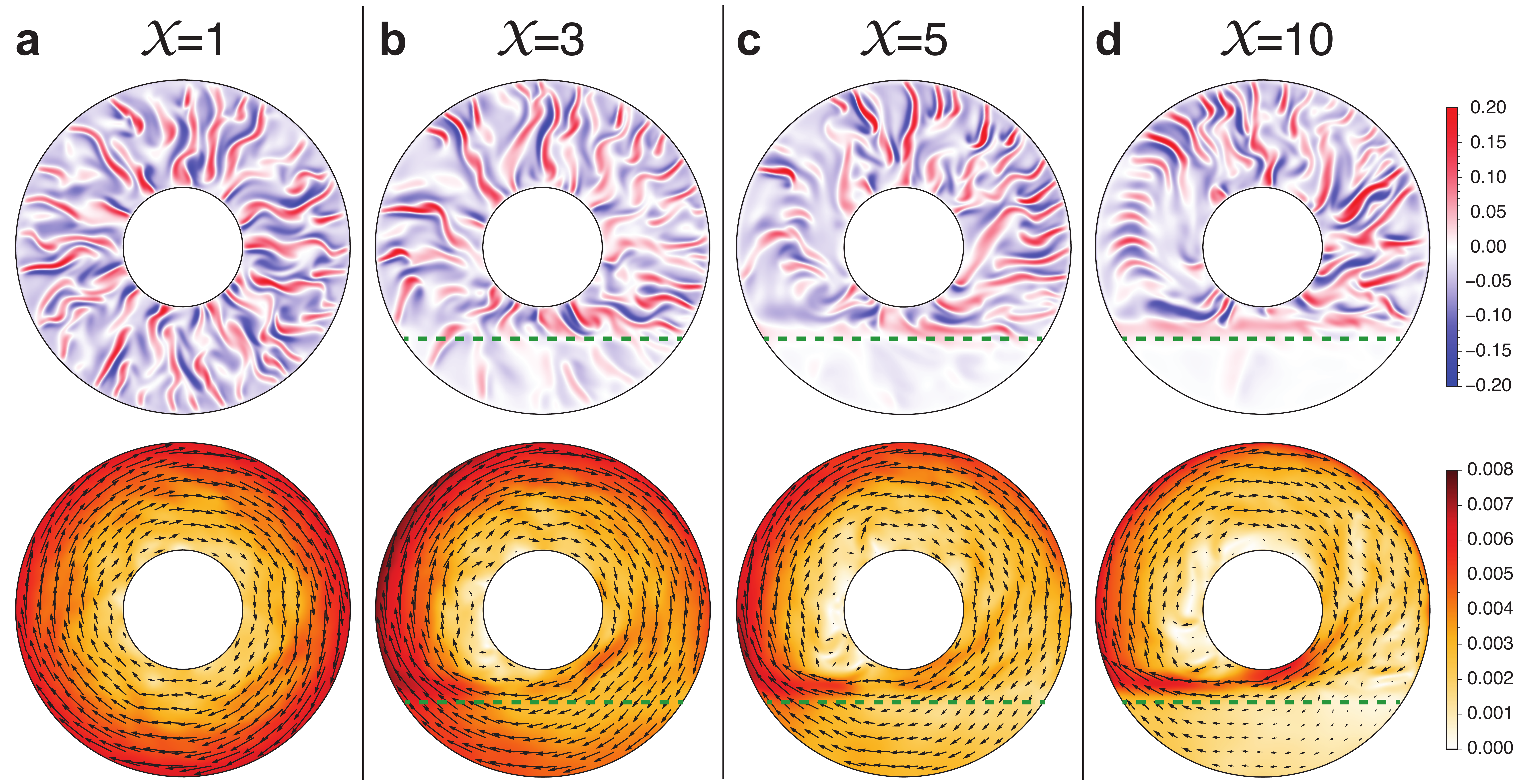} 
\caption{ {\bf Modification of core flows by a non-uniform EM drag at the CMB.}  {\bf a-d}, Snapshots of the axial vorticity (top row) and time-averaged flow maps (bottom row, colour scale indicates flow speed, arrows show direction) from our quasi-geostrophic model for $Ra=5 \cdot 10^8$, $P_m=0.1$ and different choices of ${\cal X}$.  All plots are equatorial planforms.  The Pacific region is in the bottom section of each planform, in the same location as in Fig. 2, and is delimited by a dashed green line in panels {\bf b}, {\bf c} and {\bf d}.  The colour scales on the right are common to all 4 panels. }
\end{center}
\end{figure}

\newpage

\FloatBarrier

\begin{figure}
\begin{center}
\includegraphics[width=15cm]{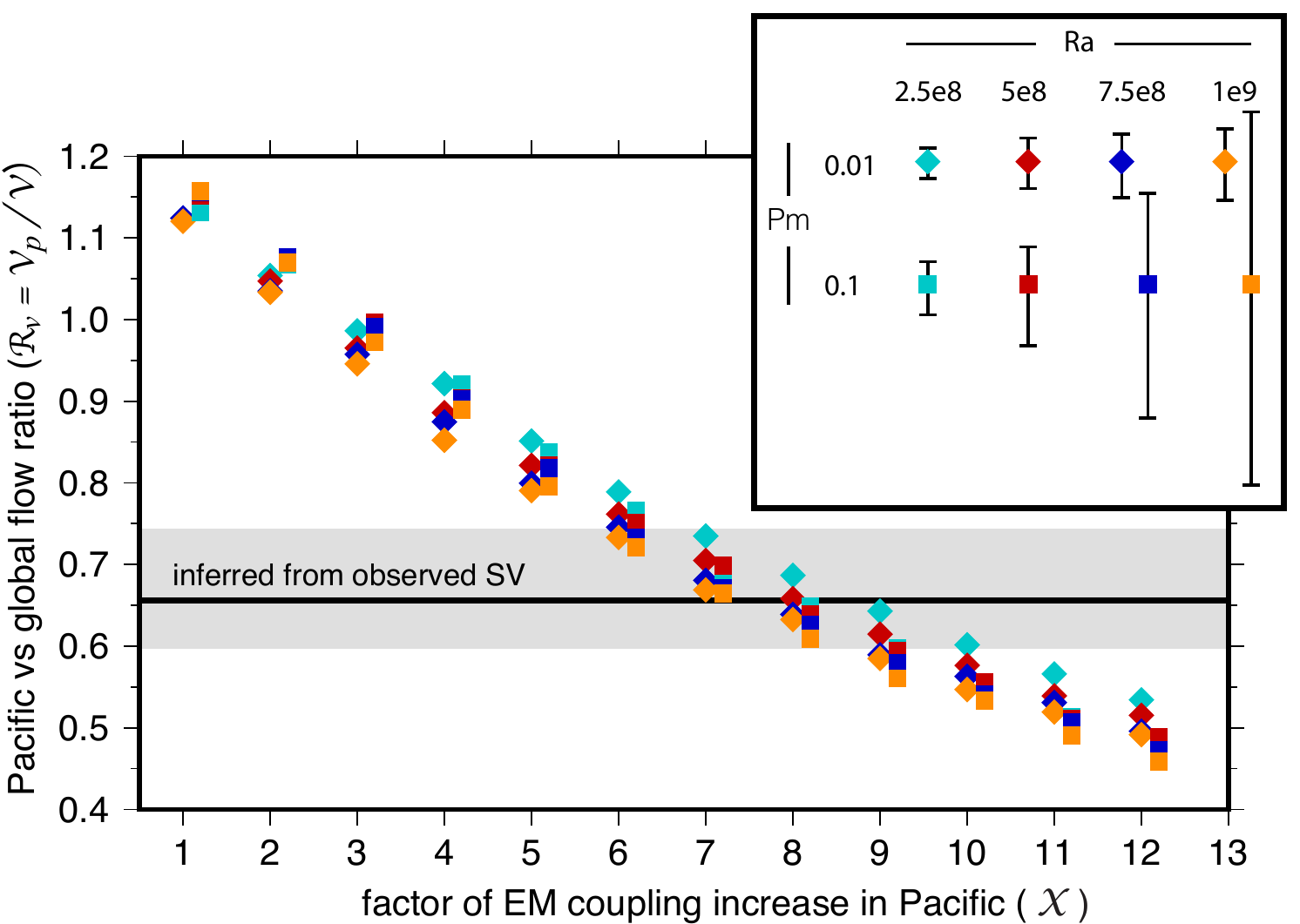} 
\caption{ {\bf Weak core flows from enhanced EM drag in the Pacific.} The ratio ${\cal R}_v={\cal V}_p/{\cal V}$ of the Pacific $({\cal V}_p)$ versus its global $({\cal V})$ r.m.s. flow speed as a function of ${\cal X}$ for values of $Ra$ and $P_m$ shown in the inset.  ${\cal R}_v$ is computed by taking the mean over a time-window of $200 \cdot\tau_A$ sampled at intervals of $0.5 \cdot\tau_A$. Error bars (inset) show the range of temporal fluctuations. Results with $P_m=0.1$ have been shifted by 0.2 in ${\cal X}$ to ease visualization.  The mean ratio ${\cal V}_p/{\cal V}=0.656$ between 1940-2010 from the model of ref. (\citen{gillet15}) is shown by the black horizontal line, with the grey band delimiting the minimum ($0.597$) and maximum ($0.744$) annual ratio.}
\end{center}
\end{figure}

\newpage

\FloatBarrier

\section*{References}

\vspace*{1cm}

\bibliographystyle{naturemag}

\begin{addendum}
 \item[Correspondence] Correspondence and requests for materials
should be addressed to Mathieu Dumberry~(email: dumberry@ualberta.ca).
\end{addendum}

\begin{addendum}
 \item This is a pdf print of an article published in Nature Geoscience. The final authenticated version is available
online at: {\blue \href{https://www.nature.com/articles/s41561-020-0589-y}{https://www.nature.com/articles/s41561-020-0589-y}}  We thank Nathana\"el Schaeffer for sharing his original numerical QG code which we extended over the course of this project and Nicolas Gillet for sharing his flow models.  Figures were created using the GMT software\cite{gmt}.  Numerical simulations were performed on computing facilities provided by WestGrid and Compute/Calcul Canada.  This work was supported by a Discovery grant from NSERC/CRSNG.  
\end{addendum}

\begin{addendum}
\item[Author Contributions] M.D. designed the project and wrote the manuscript. 
The custom numerical codes were designed and written by both M.D. and C.M.  The numerical experiments were carried and analyzed by both M.D. and C.M. 
\end{addendum}

\begin{addendum} \item[Competing Interests] The authors declare that they have no
competing financial interests.
\end{addendum}

\newpage

\begin{methods}

\section*{\large The QG magnetoconvection model}

To simulate the Earth's fluid core dynamics, we use a quasi-geostrophic (QG) numerical model of thermal convection\cite{cardin94,aubert03,gillet06} to which we add a background magnetic field and an induction equation to track the evolution of magnetic field perturbations.  The background magnetic field represents the field generated over long timescales by dynamo action.  This field is imposed and assumed steady: we conduct a magnetoconvection experiment.  The model is presented in detail in an earlier publication\cite{more18} and we give only an overview here, focusing on how we implement electromagnetic (EM) coupling at the fluid-solid boundaries.  

QG models exploit the dominance of the Coriolis acceleration in the force balance which tends to make fluid motions invariant in the direction of the rotation vector.  Only the horizontal flow components (those perpendicular to the rotation vector) need to be evolved. The flow parallel to the rotation axis contributes to vorticity generation and is not neglected: it is parameterized in terms of the horizontal flow by ensuring conservation of mass and no penetration at the spherical boundaries.  In effect, a QG model collapses a three-dimensional dynamical model within a spherical shell to a two-dimensional model on the equatorial plane.  Unlike the flow, the magnetic field inside the core is three-dimensional and cannot be approximated by an equivalent QG model. However, it is possible to capture the Lorentz force acting on the QG flows in terms of an effective horizontal magnetic field axially averaged over fluid columns\cite{more18}.  The magnetic field in our QG model represents this effective field. 

The geometry of the QG model is shown in Fig. 2 of ref. (\citen{more18}).  We restrict the solution domain to the region outside the tangent cylinder (TC, the cylinder tangent to the inner core equator).   Cylindrical coordinates $(s,\phi, z)$ are assumed with the $z$-direction aligned with the rotation axis.  The domain boundaries are $r_i$ (the radius of the TC) and $r_c$, the radius of the core-mantle boundary (CMB).  Length is scaled by $r_{c}$, time by the inverse of the angular rotational velocity $\Omega$, temperature by the superadiabatic temperature difference $\Delta T$ between the inner and outer spheres, and the magnetic field by $r_{c} \Omega \sqrt{\rho \mu}$, where $\rho$ is the reference density of the fluid (assumed uniform) and $\mu$ is the magnetic permeability of free space. 

The horizontal velocity ${\bf u}_{H}$ and magnetic field perturbation ${\bf b}_{H}$ are expanded as

\begin{equation}
    {\bf u}_H  = \overline{u_\phi} \, {\bf e}_\phi + \frac{1}{L} \bnabla \times \big( L \psi \, {\bf e}_z \big) \, ,  \hspace*{1cm}
    {\bf b}_H = \overline{b_\phi} \, {\bf e}_\phi + \frac{1}{L} \bnabla \times \big( L a \, {\bf e}_z \big) \, ,
\end{equation}
where $\psi$ and $a$ are toroidal scalars, $L = \sqrt{1 - s^{2}}$ is the half-column height, and the overbar denotes an azimuthal average; $\overline{u_\phi}$ and $\overline{b_\phi}$ capture the axisymmetric azimuthal (zonal) flow and magnetic field, respectively.   The components of ${\bf u}_H = u_s {\bf e}_s + u_\phi {\bf e}_\phi$ and the axial vorticity $\omega_z$ are defined as

\begin{subequations}
\begin{align}
    u_s & = \frac{1}{s} \frac{\partial {\psi}}{\partial \phi} \,  ,\hspace*{1cm}
        u_\phi  = \overline{u_\phi} - \left( \frac{\partial}{\partial s} + \beta \right) \psi \, , \label{eq:uh} \\ 
            \omega_z & =   \left( s \frac{\partial}{\partial s} \frac{\overline{u_\phi}}{s} + 2 \frac{\overline{u_\phi}}{s} \right)
           - \nabla_{H}^{2} \psi
           - \frac{1}{s} \frac{\partial}{\partial s} \left( s \beta \psi \right) \, .\label{eq:wz} 
\end{align}
Likewise, the components of ${\bf b}_H = b_s {\bf e}_s + b_\phi {\bf e}_\phi$ and the axial current $j_z$ are defined as

\begin{align}
    b_s &= \frac{1}{s} \frac{\partial a}{\partial \phi} \, , \hspace*{1cm}
        b_\phi  = \overline{b_\phi} - \left( \frac{\partial}{\partial s} + \beta \right) a \, , \label{eq:bh} \\  
    j_z &=   \left( s \frac{\partial}{\partial s} \frac{\overline{b_\phi}}{s} + 2 \frac{\overline{b_\phi}}{s}  \right)
           - \nabla_{H}^{2} a
           - \frac{1}{s} \frac{\partial}{\partial s} \left( s \beta a \right) \, .\label{eq:jz} 
\end{align}
\end{subequations}
Here, $\nabla_{H}^{2}=\bnabla_{H}\cdot\bnabla_{H}$ indicates the horizontal $(s,\phi)$ components of the Laplacian operator, $\bnabla_{H}$ the horizontal gradient, and the slope factor $\beta$  is a measure of how $L$ changes with $s$:

\begin{equation}
    \beta = \frac{1}{L} \frac{\partial L}{\partial s} = - \frac{s}{L^{2}} \, .
    \label{eq:beta}
\end{equation}
The $\beta$-factor ensures conservation of mass and no penetration at the spherical outer shell boundary.  The latter condition implies that spherically radial perturbations in the magnetic field should be weak compared to perturbations parallel to the boundary.  Hence, the magnetic field perturbation should approximately obey a no-penetration condition at the spherical outer shell boundary, justifying a parametrization for ${\bf b}_{H}$ of the same form as ${\bf u}_{H}$.

The model tracks the evolution of $\psi$, $a$, $\overline{u_\phi}$, $\overline{b_\phi}$ and temperature perturbations $\Theta$ through a coupled system of equations that capture the evolution of the flow, magnetic field and temperature.  The inner and outer cylindrical (non-dimensional) radii are denoted by $r_1$ and $r_2$, respectively. All variables are assumed invariant in $z$.  A steady, axially symmetric, cylindrically radial background magnetic field $B_{0s}$ is imposed; this is the only component of the background field which is dynamically important in our model.  $B_{0s}$ represents the axially averaged r.m.s. strength of the $s$-directed magnetic field $B_s$ inside the core,

\begin{equation}
B_{0s}= \sqrt{ \frac{1}{4 \pi L} \int_{-L}^{+L} \int_{0}^{2\pi}  \big(B_s\big)^2 \;  \mbox{d}\phi\;  \mbox{d}z } \, .
\end{equation}
The system of equations that captures the dynamics is:

\begin{subequations}
\begin{align}
 &   \frac{\partial \omega_z}{\partial t} + \left( u_s \frac{\partial}{\partial s} + \frac{u_\phi}{s} \frac{\partial}{\partial \phi} \right) \omega_z \, - \left(2 + \omega_z\right) \, \beta \,  u_s = - Ra^* \frac{\partial \Theta}{\partial \phi} + E \nabla_{H}^{2}  \,\omega_z \, + F_L \, ,
    \label{eq:vorz_nonaxi}\\
 &  \frac{\partial}{\partial t} \left( \frac{\overline{u_\phi}}{s} \right) + \frac{1}{s} \left( \overline{\frac{u_s}{s} \frac{\partial}{\partial s}  s u_\phi} \right) =   \frac{E}{s^{3} L} \frac{\partial}{\partial s} \left( s^{3} L \frac{\partial}{\partial s}  \left( \frac{\overline{u_\phi}}{s} \right) \right) + \Gamma_L \, ,    \label{eq:vorz_axi}  \\
&  \frac{\partial a}{\partial t} = - u_\phi  B_{0s} + \left( u_s b_\phi - u_\phi b_s \right)
           + \frac{E}{P_m} \left( \nabla_{H}^{2} \, a + \frac{2 \beta a}{s} \right) \, ,
    \label{eq:induction_nonaxi} \\
& \frac{\partial}{\partial t} \left( \frac{\overline{b_\phi}}{s} \right) =  B_{0s} \frac{\partial}{\partial s}  \left( \frac{\overline{u_\phi}}{s} \right) 
                     + \frac{1}{sL}  \frac{\partial}{\partial s}  \Big( L \left(  \overline{u_\phi b_s} - \overline{u_s b_\phi} \right)  \Big) +  \frac{E}{P_m} \frac{1}{s^{3}} \frac{\partial}{\partial s}   \left( s^{3}  \frac{\partial}{\partial s} \left( \frac{\overline{b_\phi}}{s} \right) \right)  \, ,     \label{eq:induction_axi} \\
                     & \frac{\partial \Theta}{\partial t} +\big( {\bf u}_{H} \cdot \bnabla_{H} \big) \big( T + \Theta \big) = \frac{E}{P_{r}} \nabla^{2}_{H} \Theta \, ,
    \label{eq:heat}
\end{align}
\end{subequations}
where $T$ is the conducting temperature profile 

\begin{equation}
T  = \frac{r_2}{r_2-r_1} \left(  \frac{1}{L} \ln(1+L) - 1 \right) \, . \label{eq:backT0}
\end{equation}
The non-dimensional parameters in our system are the modified Rayleigh number $Ra^* = E^{2} Ra \, P_{r}^{-1}$, the Ekman number $E = \nu \left( \Omega r_{c}^{2} \right)^{-1}$, the Prandtl number $P_{r} = \nu \kappa^{-1}$, the magnetic Prandtl number $P_{m} = \nu \eta^{-1}$ and the Rayleigh number $Ra = \alpha g_{0} \Delta T r_{c}^{3} \left( \nu \kappa \right)^{-1}$.  The parameters $\nu$, $\alpha$, $g_{0}$, $\eta$ and $\kappa$ are respectively the kinematic viscosity, thermal expansion coefficient, gravitational acceleration at $r_c$, magnetic diffusivity and thermal diffusivity.  

The $z$-component of the curl of the Lorentz force $F_L$ and the Lorentz torque $\Gamma_L$ in Eqs. (\ref{eq:vorz_nonaxi}-\ref{eq:vorz_axi}) are defined as

\begin{subequations}
\begin{align}
    F_{L} & = \left( \left( B_{0s} + b_s  \right) \frac{\partial}{\partial s} + \frac{b_\phi}{s} \frac{\partial}{\partial \phi} \right) j_z \, + {\cal F}_{em} \, ,
    \label{eq:lorentz_nonaxi} \\
    \Gamma_L & = \frac{1}{s^{3}} \frac{1}{L} \frac{\partial}{\partial s} \left( s^{3} L B_{0s} \frac{\overline{b_\phi}}{s} \right) \, + \frac{1}{s} \left( \overline{\frac{b_s}{s} \frac{\partial}{\partial s} s b_\phi} \right) \, + \overline{\cal G}_{em} \, ,\label{eq:lorentz_axi}
\end{align}
\label{eq:lorentz} 
\end{subequations}
where  ${\cal F}_{em}$  and $\overline{\cal G}_{em}$ are new additions to the model to capture EM coupling at the CMB.  Their expressions are developed in the next section.  They are

\begin{subequations}
\begin{align}
{\cal F}_{em} & = - \frac{1}{L^2} \left[ \frac{1}{s} \frac{\partial}{\partial s} ( s \, \tau_{em}^{-1} \, u_\phi ) -  \frac{1}{s} \frac{\partial}{\partial \phi} ( \tau_{em}^{-1} \, u_s ) \right] \, , \label{eq:fem}\\
 \overline{\cal G}_{em} & = - \frac{1}{s L^2} \, \overline{\tau_{em}^{-1} \,  u_\phi} \, ,\label{eq:barfem}
\end{align}
\end{subequations}
where $\tau_{em}^{-1}=1/\tau_{em}$ and $\tau_{em}$ is the characteristic attenuation timescale of a flow structure subject to EM coupling with the mantle, given by

\begin{equation}
\tau_{em}  =  \frac{E}{P_m} \frac{\sigma}{\sigma_m} \frac{1}{\Delta} \frac{1}{(B_r)^2} \label{eq:tem} \, ,
\end{equation}
where $\sigma$ and $\sigma_m$ are the electrical conductivities of the core and lowermost mantle, respectively, $\Delta$ is the (non-dimensional) thickness of the conducting layer at the bottom of the mantle, and $B_r$ is the (non-dimensional) r.m.s. radial magnetic field strength, assumed to be dominated by small length scales and taken as uniform over the CMB\cite{buffett02,koot13,buffett07}. $\tau_{em}$ is an input parameter of our model, specified as a function $s$ and $\phi$, so as to capture properly the timescale of EM attenuation in relation to the  Alfv\'en timescale  $\tau_{A}$. In practice, the term that involves $j_z$ in the definition of $F_L$ is set to zero.   The rational for this is given in ref. (\citen{more18}).  Hence, we keep only the part of the non-axisymmetric Lorentz force caused by EM coupling, and Eq. (\ref{eq:lorentz_nonaxi}) simplifies to $F_L = {\cal F}_{em}$.

At the cylindrical radial boundaries $s=r_{1,2}$ of the equatorial planform domain, we impose a fixed temperature condition ($\Theta=0$), a free-slip condition on the flow \\ $\left(\psi = \frac{\partial}{\partial s}\left(\frac{1}{s} \frac{\partial \psi}{\partial s} +\frac{\beta \psi}{s}\right) = \frac{\partial }{\partial s} \frac{\overline{u_\phi}}{s}= 0\right)$, and a vanishing non-axisymmetric magnetic potential ($a=0$). The condition on the zonal magnetic field $\overline{b_\phi}$ is constructed by integrating the induction equation over an infinitely small thickness across the radial cylindrical boundary, where $\overline{b_\phi}$ is further assumed to be continuous.  This gives a condition relating the discontinuity in $\overline{u_\phi}$ to the discontinuity in the $s$-gradient of $\overline{b_\phi}$.  The condition at $s=r_2$ is

\begin{subequations}
\begin{equation}
 \frac{E}{P_m} \left[ \frac{\partial}{\partial s} \left( \frac{\overline{b_\phi}}{s} \right) + 2 \left( \frac{\overline{b_\phi}}{s^2}  \right) \right] \,  + \left( B_{0s} \right)^2 \, \overline{\tau_{em}} \left( \frac{\overline{b_\phi}}{s}\right) =  - B_{0s} \left( \frac{\overline{u_\phi}}{s} \right) \, , \label{eq:bcbphis2}
\end{equation}
where $\overline{\tau_{em}}$ is the axisymmetric part of Eq. (\ref{eq:tem}) and where we have used $B_r = B_{0s}$.  This condition is consistent with our formulation of EM coupling at the CMB, and is based on the assumption that the conductivity in the lowermost mantle is concentrated in a thin layer of thickness $\Delta \ll 1$ within which we can use the following approximation,  

\begin{equation}
\frac{\partial}{\partial s} \left( \frac{\overline{b_\phi}}{s}  \right) = - \frac{1}{\Delta}  \left( \frac{\overline{b_\phi}}{s} \right) \, .\end{equation}
The condition at $s=r_1$ (TC) is 
\begin{equation}
 2 \frac{E}{P_m} \left[ \frac{\partial}{\partial s} \left( \frac{\overline{b_\phi}}{s} \right) \right] =  - B_{0s} \left[ \left( \frac{\overline{u_\phi}}{s} \right)  - \overline{\omega_{tc}} \right] \, , \label{eq:bcbphis1} 
\end{equation} 
\end{subequations}
where we have assumed equal values of $P_m$ and a symmetric $s$-derivative of $\overline{b_\phi}$ on either side of the TC.  $\overline{\omega_{tc}}$ is the angular velocity of the TC which is imposed to capture the time-average angular momentum balance of the inner core - fluid core - mantle system (see below).  

Compared to a three-dimensional model self-generating a dynamo, an advantage of our QG model is that we capture correctly the ratio of convective velocities to the speed of propagation of Alfv\'en waves, a ratio known as the Alfv\'en number which is approximately equal to 0.01 in the Earth's core.  This is especially important for modelling the decadal timescale dynamics\cite{more18}, including the effects of EM drag on the flow.  Flows in Earth's core that vary on 10-100 yr timescale are expected to be rigid\cite{jault08}, so our QG model is tailored to capture their dynamics and is a good analog for Earth. 

\section*{\large Electromagnetic coupling}

A differential tangential motion between the core and an electrically conducting lowermost mantle at the CMB shears the local radial magnetic field $B_r$.  This creates a tangential magnetic field which, interacting with $B_r$, leads to a tangential EM stress\cite{rochester60} by the fluid core on the mantle (and vice-versa), a 'magnetic friction' which acts to slow down core flows\cite{dumberry08c,gillet17}.  

Following Braginsky\cite{braginsky70}, we assume that the (dimensional) electrical conductivity of the mantle $\sigma'_m$ is high within a layer of (dimensional) thickness $\Delta'$ just above the CMB. $\Delta'$ may vary geographically but it is assumed to be everywhere much thinner than the magnetic skin depth $\delta_m = \sqrt{2/\omega' \mu\sigma'_m}$, where $\omega'$ is the characteristic (dimensional) frequency of flow fluctuations.  We calculate the magnetic perturbation in this layer and its contributions to the $z$-integrated equations for the vorticity (\ref{eq:vorz_nonaxi}) and zonal flow (\ref{eq:vorz_axi}), captured respectively by the terms ${{\cal F}}_{em}$ and $\overline{{\cal G}}_{em}$ in Eqs. (\ref{eq:lorentz_nonaxi}-\ref{eq:lorentz_axi}).  $\overline{{\cal G}}_{em}$ is given by

\begin{subequations}
\begin{equation}
\overline{{\cal G}}_{em} =   \frac{B_r \, \overline{b_\phi}^{(cmb)}}{s L^2} \, ,\label{eq:barfem2}
 \end{equation}
where $\overline{b_\phi}^{(cmb)}$ is the axisymmetric azimuthal field perturbation induced by differential core-mantle motion.  The superscript $cmb$ emphasizes that it is distinct from $\overline{b_\phi}$, the azimuthal field axially averaged over the whole fluid column.  When assuming a uniform $B_r$, ${{\cal F}}_{em}$ is given by

\begin{equation}
{{\cal F}}_{em} =   \frac{B_r \, j^{(cmb)}_z}{L^2} =  \frac{1}{L^2} \left( \frac{1}{s} \frac{\partial}{\partial s} s B_r b^{(cmb)}_\phi  -\frac{1}{s} \frac{\partial}{\partial \phi} B_r b^{(cmb)}_s \right) \, , \label{eq:fem2}
 \end{equation}
\end{subequations}
where $j^{(cmb)}_z$ and $b^{(cmb)}_{s,\phi}$ are the non-axisymmetric axial current and $s,\phi$ magnetic field components at the CMB induced by differential core-mantle motion. 

The (dimensional) magnetic field perturbation vector tangential to the spherical surface at the CMB is \cite{buffett92,buffett98}

\begin{subequations}
\begin{equation}
 {\bf b}_H^{(cmb)} = -\mu \sigma'_m  B_{r} \Delta' \, {\bf u}_H^{(cmb)} \, .
 \end{equation}
Expressed in non-dimensional form, and in terms of the QG flow components, this gives

\begin{align}
\overline{b_\phi}^{(cmb)} & = - \frac{P_m}{E} \frac{1}{\sigma} \, \overline{ \sigma_m  \Delta B_{r} {u_\phi}}= - \overline{ \left(  \frac{u_\phi}{B_{r} \, \tau_{em} } \right) }\, , \label{eq:barbcmb}\\
 b_{s,\phi}^{(cmb)} & = - \frac{P_m}{E} \frac{\sigma_m}{\sigma}\,  \Delta  B_{r} \,  u_{s, \phi} = - \frac{u_{s,\phi}}{B_r \, \tau_{em} }  \, . \label{eq:bcmb}
 \end{align}
 \end{subequations}
Inserting these expressions into Eqs. (\ref{eq:barfem2}-\ref{eq:fem2}) gives the expressions of ${\cal F}_{em}$ and $\overline{\cal G}_{em}$ in Eqs.  (\ref{eq:fem}-\ref{eq:barfem}).  Because of the symmetry about the equatorial plane that is built into the QG model, it is the averaged EM coupling between the two hemisphere that QG flows are sensitive to.

\section*{\large Axial angular momentum balance}

The mean westward flow at the CMB observed in Fig. 2 of the main text leads to a mean westward EM torque on the mantle.  An equal and opposite eastward torque is applied by the mantle on the fluid core.  On a time average, the total torque on the fluid core must vanish, so a westward torque must be applied on the fluid core at the ICB.  In the Earth's core, this is achieved by a thermal wind flow structure inside the TC which involves a mean eastward flow at the ICB\cite{aurnou96,olson99,pichon16}.  This flow exerts an eastward EM torque on the inner core, and thus a reversed westward torque on the fluid core. 

The solution domain of our QG model is restricted to the region outside the TC.  We must therefore substitute the westward torque on the fluid core at the ICB by an equivalent westward torque at the TC.  The equation governing the total axial angular momentum of the fluid core in our system is obtained by multiplying Eq. (\ref{eq:vorz_axi}) by $ s^3 L$ and integrating from $s=r_1$ to $r_2$.  With our choice of boundary conditions, the Reynolds stress term, the non-linear Lorentz torque and the viscous torque all vanish\cite{more18}, leaving 

\begin{subequations}
\begin{equation}
\int_{r_1}^{r_2}  s^3 L \frac{\partial}{\partial t} \left( \frac{\overline{u_\phi}}{s} \right)  \, ds =   -  \int_{r_1}^{r_2}  \frac{s^2}{L} \, \overline{\tau_{em}^{-1} \,  u_\phi} \, ds \, + \Big[ s^2 L B_{0s} \overline{b_\phi} \Big]^{r_2}_{r_1}  \, .
\label{eq:focangmom}
\end{equation}
Since $L\rightarrow 0$ as $s \rightarrow r_2$, the last term is dominated by the contribution at the TC ($s=r_1$). The steady state angular momentum balance is then 

\begin{equation}
  -  \int_{r_1}^{r_2}  \frac{s^2}{L} \, \overline{\tau_{em}^{-1} \,  u_\phi} \, ds \, - \Big[ s^2 L B_{0s} \overline{b_\phi} \Big]_{tc}  = 0 \, .
\label{eq:focangmom2}
\end{equation}
The first term represents the mean torque that the mantle exerts on the fluid core by EM coupling.   The second term is the mean EM torque exerted on the fluid core at the TC. 

Our QG model is designed such that, as is the case for Earth, a steady-state dominantly westward flow ($\overline{u_\phi} <0$) is present in a large portion of the fluid core.  In order to achieve this, the EM torque exerted by the TC must be negative, or westward.  We prescribe this torque by assuming that the region inside the TC has the same electrical conductivity as the fluid core and is differentially rotating with angular velocity $\overline{\omega_{tc}}$.  By shearing the $B_{0s}$ magnetic field, this differential rotation leads to a cusp in $\overline{b_\phi}$ at the TC specified by the boundary condition of Eq. (\ref{eq:bcbphis1}).   Writing ${\partial \overline{b_\phi}}/{\partial s} = - {\overline{b_\phi}}/{\delta}$, where $\delta$ is a characteristic length scale, $\overline{b_\phi}$ at the TC is

\begin{equation}
\overline{b_\phi} \big|_{tc} = 2 \delta B_{0s}  \frac{P_m}{E} \Big( \overline{u_\phi} \big|_{tc}  - r_1 \overline{\omega_{tc}} \Big)  \, .
\end{equation}
Inserting this in the torque balance of Eq. (\ref{eq:focangmom2}), the latter becomes
\begin{equation}
  -  \int_{r_1}^{r_2}  \frac{s^2}{L} \, \overline{\tau_{em}^{-1} \,  u_\phi} \, ds \, = 2 \delta \frac{P_m}{E} \left[ s^2 L  B_{0s}^2 \right]_{tc} \Big( \overline{u_\phi} \big|_{tc}  - r_1 \overline{\omega_{tc}} \Big) \, .
\label{eq:focangmom3}
\end{equation}
\end{subequations}
By imposing a westward rotation of the TC ($\overline{\omega_{tc}} < 0$), the net EM torque by the TC is westward.  The mean flow $\overline{u_\phi}$ must be then be generally westward such that the global EM torque by the mantle (left-hand side of Eq. \ref{eq:focangmom3}) is eastward and angular momentum equilibrium of the fluid core is achieved.  

 \section*{\large Parameters, conductance model and numerical implementation}
 
The cylindrical radial boundaries of our domain of integration are $r_1=0.35$ and $r_2=0.98$.  Limiting our domain to $r_2= 0.98$ instead of $r_2 = 1$ is convenient, as it allows us to use a slightly coarser grid space and longer timesteps (since $\beta \rightarrow -\infty$ as $s \rightarrow 1$), but the solutions are otherwise not altered by this choice.  All of our model calculations use the following common set of parameters: $E=5 \cdot 10^{-6}$, $P_r=1$ and $\overline{\omega_{tc}}=-0.005$.  We have used four different choices of Rayleigh numbers ($2.5\times10^8$, $5\times10^8$,  $7.5\times10^8$ and $10^9$) and two different values of magnetic Prandtl numbers ($0.1$ and $0.01$).  In all our numerical experiments, $B_{0s}$ varies with $s$ and is specified by the function 
\begin{equation}      
B_{0s} = 0.15 \cdot \bigg[ 1- \frac{4}{5} \cdot \exp \Big( - 10 \cdot (r_2-s) \Big) \bigg] \, . \label{eq:bos}
\end{equation}
The amplitude of $B_{0s}$ sets the speed of Alfv\'en waves, and with the selected amplitude, the Alfv\'en number for all the simulations shown in Fig. 4 is between 0.02 and 0.03, similar to that expected in Earth's core\cite{more18}. We note that choosing different forms of $B_{0s}$, while keeping the same amplitude, does not substantially change the patterns of ${\bf u}_H$ and ${\bf b}_H$ in our model.  

The projection of the boundary of an equatorial spherical cap onto the equatorial planform is a linear segment. Using cartesian coordinates where $y = s \cdot \sin \phi$, we define the Pacific region to be in the $y<0$ quadrant.  We build a smoothly varying function of the EM attenuation timescale $\tau_{em}(y)$ as a function of $y$, reaching a minimum value $\tau_{em}^p$ in the Pacific cap, and a maximum value of $\tau_{em}$ away from it.  Defining $y_o$ ($<0$) as the segment delimiting the Pacific cap, $\tau_{em}(y)$ is specified by 

\begin{subequations}
\begin{align}
\mbox{for $y \le y_o$:} \hspace*{1cm} & \frac{1}{\tau_{em}(y)} =  \frac{\cal X}{\tau_{em}} \, ,  \\ 
\mbox{for $y > y_o$:}  \hspace*{1cm}  & \frac{1}{\tau_{em}(y)} =  \frac{1}{\tau_{em}} \left( 1 + ({\cal X} - 1) \bigg[  \exp \Big( - \big(\lambda \cdot (y- y_o)\big)^{\gamma} \Big) \bigg] \right) \, ,
\end{align}
\end{subequations}
where ${\cal X} = \tau_{em}/\tau_{em}^p$ is the factor of increase of EM coupling strength in the Pacific cap.  For all model runs, we have used $y_o=-0.7$, $\lambda=5$ and $\gamma=4$. 

The dimensional EM timescale of attenuation is approximately $\tau'_{em} =30$ yr for a conductance of $10^8$ S (ref. \citen{dumberry08c}), or equivalent to $5\cdot \tau'_A$ for an Alfv\'en timescale of $\tau'_A = 6$ yr (ref. \citen{gillet10}).  In our model, the (non-dimensional) Alfv\'en time is given by $\tau_A = 2\cdot(r_2-r_1)/ \langle B_{0s} \rangle$, where $\langle B_{0s} \rangle$ is the mean $B_{0s}$ between $r_1$ and $r_2$, and it is approximately $\tau_A \approx 9$.  Hence, by choosing $\tau_{em} = 5 \cdot \tau_A = 45$, we ensure that the ratio of $\tau_{em}/ \tau_A$ in our model is the same as that in Earth's core. Within the Pacific cap, $\tau_{em}$ is smaller by a factor ${\cal X}$.  We have used a range of ${\cal X}$ from 1 to 12.   For the largest ${\cal X}$ value, given that the Pacific cap is approximately 25\% of the surface of the CMB, the globally averaged EM attenuation timescale is reduced by a factor 3.2, equivalent to a global conductance of $3.2 \cdot 10^8$ S, which is still broadly compatible with observations\cite{buffett92,holme98b,schaeffer16b}.

The equations of our model are solved by a semi-spectral method, where variables are defined at discretized radial points specified by a Chebychev grid and expanded in Fourier modes in the azimuthal direction.  For all model runs, we have used 301 radial points and 256 Fourier modes. The resulting discrete equations are evolved using a combination of a Crank-Nicolson method for the linear terms and a second-order Adams-Bashforth scheme for the non-linear terms, using fixed time-steps of $dt=0.0005$.

\end{methods}

\begin{addendum}
 \item[Data Availability] The datasets generated as part of this study, together with the GMT scripts and data files necessary to reproduce all figures are freely accessible on UAL Dataverse at \\ {\blue \href{https://doi.org/10.7939/DVN/TL8BP6}{https://doi.org/10.7939/DVN/TL8BP6}}.
\end{addendum}

\begin{addendum}
 \item[Code Availability] All source codes used to generate the numerical simulations presented in this work are freely accessible on UAL Dataverse at {\blue \href{https://doi.org/10.7939/DVN/TL8BP6}{https://doi.org/10.7939/DVN/TL8BP6}}.
\end{addendum}

\newpage

\section*{Additional References}

\vspace*{1cm}

\bibliographystyle{naturemag}

\end{document}